\documentclass[11pt,a4paper]{article} 

\usepackage{jheppub}
\usepackage[T1]{fontenc} 
\usepackage[utf8]{inputenc}

\newcounter{MBQ}

\newcommand\MSB{\ensuremath{\overline{\rm MS}}}

\newcommand\mB{\ensuremath{\overline{m}}}
\newcommand\mr{\ensuremath{m}}
\newcommand\mum{\ensuremath{\mu_m}}
\newcommand\mur{\ensuremath{\mu}}

\newcommand\mpole{\ensuremath{m_{\scriptscriptstyle P}}}
\newcommand\mathd{{\rm d}}

\newcommand\as{\ensuremath{\alpha_s}}
\newcommand\lambdaQCD{\ensuremath{\Lambda_{\scriptscriptstyle \rm QCD}}}
\newcommand\LambdaQCD{\lambdaQCD}

\newcommand\nc{n_{\rm c}}
\newcommand\nl{n_l}
\newcommand\cf{C_F}
\newcommand\ac{c^{(\rm as)}}
\newcommand\acb{\tilde{c}^{(\rm as)}}

\begin{document}
\allowdisplaybreaks

\begin{titlepage}

\begin{flushright}
DESY~16-082\\
TTP16-016\\
TUM-HEP-1044/16\\
arXiv:1605.03609 [hep-ph]\\[0.0cm]
June 02, 2017

\end{flushright}

\vskip1cm
\begin{center}
{\fontsize{15}{19}\bf\boldmath
On the ultimate uncertainty of the top quark pole mass}
\end{center}

\vspace{0.8cm}
\begin{center}
{\sc M.~Beneke$^{a}$, P.~Marquard$^{b}$, P.~Nason$^{c}$, 
M.~Steinhauser$^{d}$}\\[6mm]
{\it ${}^a$Physik Department T31,\\
James-Franck-Stra\ss e, 
Technische Universit\"at M\"unchen,\\
D--85748 Garching, Germany\\
\vspace{0.3cm}
${}^b$Deutsches Elektronen Synchrotron DESY,\\
Platanenallee 6, D--15738 Zeuthen, Germany\\
\vspace{0.3cm}
${}^c$INFN, Sezione di Milano Bicocca, 20126 Milan, Italy\\
\vspace{0.3cm}
${}^d$ Institut f{\"u}r Theoretische Teilchenphysik,\\ 
Karlsruhe  Institute of Technology (KIT),\\ 
D--76128 Karlsruhe, Germany}
\\[0.3cm]
\end{center}


\vspace{0.7cm}
\begin{center} {\bf Abstract} \end{center}
\begin{abstract}
\noindent
We combine the known asymptotic behaviour of the QCD perturbation 
series expansion, which relates the pole mass of a heavy quark to the 
\MSB{} mass, with the exact series coefficients up to 
the four-loop order to determine the ultimate uncertainty of the 
top-quark pole mass due to the renormalon divergence. We perform 
extensive tests of our procedure by varying the number of colours 
and flavours, as well as the scale of the strong coupling and the 
\MSB{} mass. Including an estimate of the internal bottom and charm 
quark mass effect, we conclude that this uncertainty is 
around 110~MeV. We further estimate the additional 
contribution to the mass relation from the five-loop correction and 
beyond to be around 300~MeV. 
\end{abstract}
\end{titlepage}



\section{Introduction}

The top quark mass is a fundamental parameter of the Standard Model (SM). 
Due to its large size, it has non-negligible impact in the precision tests 
of the SM. After the discovery of the Higgs boson and the measurement of 
its mass, the values of the $W$ and top mass are strongly correlated, such 
that a precise determination of both parameters would lead to a SM test of 
unprecedented precision~\cite{Agashe:2014kda}. Indeed, there is presently 
some tension between the value of the top mass $177\pm 2.1$~GeV fitted from 
electroweak data and from its direct measurement~\cite{Agashe:2014kda}, 
for which the combination of the Tevatron and LHC data yields the
1.6~$\sigma$ lower value of  $173.34\pm 0.27\pm 
0.71$~GeV~\cite{ATLAS:2014wva}. The value of the top mass is also crucial to 
the issue of stability of the SM 
vacuum (see~\cite{Buttazzo:2013uya} for a recent analysis).
The Higgs quartic coupling decreases at high scales, eventually becoming 
negative. This evolution is very sensitive to the top mass value. For 
example, a top mass near 171~GeV would imply that the quartic coupling may
vanish at the Planck scale, rather than turn negative.

The standard direct determination of the top mass at hadron colliders, being 
based upon observables that are related to the mass of the system comprising 
the top decay products, are quoted as measurements of the pole mass. On the 
other hand, it seems more natural to use the \MSB{} mass in both precision 
electroweak observables and in vacuum stability studies.  
In~\cite{Marquard:2015qpa} the relation between the \MSB{} and pole mass for 
a heavy quark (the ``mass conversion formula'' from now on) has been computed 
to the fourth order in the strong coupling $\alpha_s$. Assuming the value of 
$163.643$~GeV for the top-quark \MSB{} mass 
$\overline{m}_t=m_t(\overline{m_t})$, and assuming 
$\as^{(6)}(m_t)=0.1088$, we have~\cite{Marquard:2015qpa}
\begin{equation}\label{eq:marquard2}
\mpole = 163.643+7.557+1.617+0.501 + (0.195\pm 0.005)\,{\rm GeV}
\end{equation}
for the series expansion of the mass conversion formula. The last term 
from the fourth order correction is less than one half of the 
third order one.

It is also known that the mass conversion formula is affected 
by infrared (IR) renormalons~\cite{Beneke:1994sw,Bigi:1994em,Beneke:1994rs}. 
This means that there are factorially growing terms of infrared origin in the 
perturbative expansion, such that the expansion starts to diverge at some 
order. If the series is treated as an asymptotic expansion, the ambiguity
in its resummation is of order of a typical hadronic scale.
Because of this, it is often stated that the ultimate 
accuracy of top pole mass cannot be below a few hundred MeV.
One of the goals of this work is to make this estimate more precise.  

It is remarkable that the perturbative relation between the pole and \MSB{} 
mass of a heavy quark appears to be dominated by the leading infrared 
renormalon already in low orders~\cite{Beneke:1994qe,Ball:1995ni}. This 
observation was used in previous 
work \cite{Pineda:2001zq, Lee:2003hh,Hoang:2008yj}, and 
more recently in \cite{Ayala:2014yxa,Lee:2015owa} to estimate the 
unknown normalization of the leading IR renormalon, and mostly applied 
in the context of bottom physics. In the context of top physics, the 
importance of this issue was raised recently in \cite{Nason:2016tiy}. 
The purpose of this work is to combine 
the newly available four-loop coefficient \cite{Marquard:2015qpa} 
in the mass conversion formula with the known structure of the first 
infrared renormalon singularity~\cite{Beneke:1994rs} to determine the 
normalization constant and discuss its impact on top physics. We also 
perform an analysis of the dependence on the number of colours 
and flavours, which is by itself of interest, and stability tests with 
respect to variations of the scale of the strong coupling and 
$\overline{\rm MS}$ mass. This leads to an expression for the 
mass conversion factor including an estimate of the contributions 
beyond four loops, and an estimate of the irreducible error.


\section{Reminder}
\label{sec:reminder}

The renormalon divergence is a manifestation of the fact that the 
mass conversion formula, while infrared finite is sensitive to small 
loop momentum. In the case of the pole mass this sensitivity is 
particularly strong, namely linear, resulting in rapid divergence 
of the perturbative expansion, and an infrared sensitivity of order 
$\lambdaQCD$~\cite{Beneke:1994sw,Bigi:1994em}. The ambiguity in defining 
the pole mass is therefore of similar size. This is not surprising 
as the pole mass of a quark is not an observable due to confinement 
and the difference with the physical heavy meson masses is also of 
order $\lambdaQCD$. Unlike other heavy quarks, the top quark decays 
on hadronic time scales, and thus the propagator pole position acquires 
an imaginary part. The renormalon divergence is not 
altered \cite{Smith:1996xz} by the fact that the top quark is unstable  
with a width larger than $\lambdaQCD$ and hence does not form bound states. 
The finite width simplifies the perturbative treatment of top quarks, 
since it provides a natural IR cut-off, and there exists no quantity for 
which the pole mass would ever be relevant. But the infrared sensitivity 
of the QCD corrections to the mass conversion factor, which causes the 
divergence,  remains unaffected by the width. 

Slightly more technically, the divergence arises from logarithmic 
enhancements of the loop integrand. Heuristically, this can be 
understood by noticing that the running coupling evaluated at the scale 
$l$ of the loop momentum has the expansion
\begin{equation}
\as(l)=\frac{1}{b_0 \ln \,l^2/\Lambda^2_{\rm QCD}}=
\frac{\as(m)}{1-\as(m) b_0 \ln m^2/l^2}
= \sum_1^\infty \as^n(m) \,b_0^n \ln^n \frac{m^2}{l^2}\,.
\end{equation}
The IR contribution to the last loop integration in the 
$(n+1)$-loop order then takes the form 
\begin{equation}
\label{eq:selfenexp}
\delta m^{(n+1)} 
\propto \as^{n+1}(m) \int^m \!\mathd l\,b_0^n
\ln^n \frac{m^2}{l^2}
= m \left(2 b_0\right)^n \as^{n+1}(m)\,n!\,.
\end{equation}
With this behaviour the series of mass corrections reaches a minimal 
term of order 
\begin{eqnarray}
m\,(2 b_0)^n \as^{n+1} n! &\approx& m\,\as\, 
n^{-n} \,(\sqrt{2\pi} n^{n+1/2} e^{-n}) 
\approx
m\,\sqrt{\frac{\pi\alpha_s}{b_0}}\,
\exp\left(-\frac{1}{2 b_0 \as}\right) 
\nonumber\\ 
&\approx&\sqrt{\frac{\pi\alpha_s}{b_0}}\,\LambdaQCD,
\label{eq:minestimate}
\end{eqnarray}
when  $n \approx 1/(2 b_0 \as)$ and then diverges.
Asymptotic expansions can sometimes be summed using the Borel transform. 
Given a power series
\begin{equation}
f(\as)=\sum_{n=1}^{\infty} c_n \as^n\,,
\end{equation}
the corresponding Borel transform is defined by
\begin{equation}
  B[f](t)=\sum_{n=0}^{\infty} c_{n+1}\,\frac{t^n}{n!}\,.
\end{equation}
The Borel integral 
\begin{equation}
\int_0^\infty dt\,e^{-t/\as}\, B[f](t)
\end{equation}
has the same series expansion as $f(\as)$ and provides the exact result 
under suitable conditions. However, for the case of (\ref{eq:selfenexp}), 
where $c_{n+1} = (2b_0)^n n!$, the Borel integral 
\begin{equation}
\int_0^\infty dt\, e^{-t/\as}\,\frac{1}{1-2b_0 t}
\label{eq:simpleborel}
\end{equation}
cannot be performed because of the pole at $t=1/(2 b_0)$. 
We can introduce some prescription for handling the pole in the integral, 
as, for example, the principal value prescription. 
Whether or not this reconstructs the exact result, an ambiguity
remains, quantified by the imaginary part of the integral when going above 
or below the singular point. A commonly used procedure is
to define this ambiguity to be equal to the imaginary part of the
integral divided by Pi (see, e.g.,~\cite{Beneke:1998ui}, 
section~5.2). For (\ref{eq:simpleborel}), this yields 
\begin{equation}
\Lambda_{\rm QCD}/{(2 b_0)}\,.
\label{eq:boralambiguity}
\end{equation}
In the range of $\as$ values considered in this paper, the ambiguity is 
close to the size of the smallest term in 
(\ref{eq:minestimate}).\footnote{Note, however, the different 
parametric dependence on $\alpha_s$ of (\ref{eq:minestimate}) and 
(\ref{eq:boralambiguity}). The correct dependence is that of 
(\ref{eq:boralambiguity}), for the following reason: The
typical width of the region where the minimal term is attained grows 
parametrically as $\sqrt{1/(2 b_0 \as)}$. The accuracy of an asymptotic series 
is better estimated by the minimal term times the factor accounting for the 
number of terms in this region, 
which makes (\ref{eq:minestimate}) parametrically consistent
with (\ref{eq:boralambiguity}). Numerically, this factor turns out to 
be of order one for the applications considered in this paper, as
will be confirmed in section~\ref{sec:extrapolation} below. In case of 
doubt, the estimate from the ambiguity of the Borel integral should 
be the preferred choice.} 

It can be shown \cite{Beneke:1994rs} that while the precise asymptotic 
behaviour of the mass conversion formula differs from the simple ansatz 
employed in this section for illustration, as discussed below, the 
ambiguity is exactly proportional to $\lambdaQCD$, which evaluates 
to about $250~$MeV in the $\overline{\rm MS}$ scheme. In the 
remainder of this work, we aim to quantify the 
proportionality factor.


\section{The leading pole mass renormalon}
\label{sec:renormalon}

We write the perturbative expansion of the mass conversion formula 
as
\begin{eqnarray}
\mpole &=& \mr(\mum) \bigg(1+\sum_{n=1}^\infty c_{n}(\mu,\mum,m(\mum)) 
\,\alpha_s^{n}(\mu) \bigg)   \,.
\label{eq:beneke2} 
\end{eqnarray}
Here $\as(\mu)$ is the $\overline{\rm MS}$ coupling in the $n_l$ light 
flavours theory, and 
$\mr(\mum)$ stands for the \MSB{} mass evaluated at the scale 
$\mum$. (In the following we will consider different scale choices for the 
heavy quark mass and the strong coupling constant). We also use $\mB$ to 
denote the \MSB{} mass evaluated self-consistently at a scale equal to the 
mass itself, i.e.
\begin{equation}
\mB=\mr(\mB).
\end{equation}

The leading IR renormalon divergence implies the following large-$n$  
behaviour of the perturbative coefficients~\cite{Beneke:1994rs} 
(and \cite{Beneke:1998ui}\footnote{
The perturbative coefficients $r_n$ in this reference are related to 
those employed here by $r_n = c_{n+1}$. With this notation the number of 
loops contributing to $c_n$ is $n$.}, eq.~(5.90))
\begin{eqnarray}
\label{eq:cnasymp}
c_{n}(\mu,\mu_m,\mr(\mum)) & \underset{n\to\infty}\longrightarrow & 
N \ac_{n}(\mu,\mr(\mum)) \equiv 
N \frac{\mu}{\mr(\mum)}\,\acb_{n}\,, 
\end{eqnarray}
where
\begin{eqnarray}
\label{eq:ctildenasymp}
\tilde{c}_{n+1}^{(\rm as)} &=& 
(2 b_0)^n\, \frac{\Gamma(n+1+b)}{\Gamma(1+b)} 
\left(1+\frac{s_1}{n+b}+\frac{s_2}{(n+b)(n+b-1)}
+\cdots \right).
\end{eqnarray}
It is remarkable that $b=b_1/(2 b_0^2)$ and the $s_i$ coefficients of the 
sub-leading ${\cal O}(1/n^i)$ behaviour can all be given in terms of the 
coefficients of the beta-function~\cite{Beneke:1994rs}. 
The relevant expressions are collected in 
appendix~\ref{sec:appendixA}. We also note that the scale $\mum$ at which 
$m$ is evaluated does not appear explicitly on the right-hand side of  
(\ref{eq:cnasymp}) and hence is irrelevant in (\ref{eq:beneke2}) as far 
as the large-$n$ behaviour is concerned. The dependence on the scale 
$\mu$ of the strong coupling is compensated by the factor $\mu$ in 
front of $\acb_{n+1}$ in~(\ref{eq:cnasymp}).
With these definitions the normalization $N$ is independent of 
$\mu$ and $\mum$. It cannot however be computed rigorously with present 
perturbative techniques in general, but in the limit of large negative or 
positive $\nl$ it assumes the value~\cite{Beneke:1994sw}
\begin{equation}\label{eq:Nlargenl}
\lim_{|\nl| \to \infty} N=\frac{\cf}{\pi} \times e^{\frac{5}{6}}\,,
\end{equation}
which equals  $0.97656$ for $\nc=3$ ($\cf=4/3$).

In the following we compare the exactly known low-order coefficients of the 
perturbative expansion in the mass conversion relation with their expected 
asymptotic behaviour. By definition (see (\ref{eq:cnasymp})) the 
normalization $N$ is given by 
\begin{eqnarray}
\label{eq:Ndet}
N&=&\lim_{n\to\infty} \frac{c_{n}(\mu,\mum,\mr(\mum))}
{\ac_n(\mu,\mr(\mum))}\,.
\end{eqnarray}
We now determine $N$ by evaluating the above expression for $n=1,2,3,4$, 
for which $c_{n}(\mu,\mum,\mr(\mum))$ is known. 
To this end the result of~\cite{Marquard:2015qpa} for the four-loop 
coefficient has been expressed in terms of the strong coupling
constant with $\nl$ flavours rather than $\nl+1$, since the asymptotic 
expression refers to the $\nl$ massless flavour theory. We also use 
unpublished results \cite{unpublished} for the $n_l$, $\nc$, $\mu$ and 
$\mum$ dependence of the four-loop coefficient.
In addition to the ratio $c_n/\ac_{n}$ for $n$ from 1 to 4 
we consider the relative difference between the $N$ estimates performed 
using the third and the fourth order coefficients, defined as
\begin{equation}
  \Delta_{34}=2\,\frac{|c_3/\ac_3-c_4/\ac_4|}{|c_3/\ac_3+c_4/\ac_4|}.
\end{equation}
The value of $\Delta_{34}$ can be considered to be an estimate of how
close is the third order coefficient to the asymptotic value. 
It is likely to be an overestimate of the deviation of the fourth order
coefficient from the asymptotic formula and should not be taken as an 
error on the normalization $N$.

We report our results in table \ref{tab:nc3vsnf} for $\mum=\mB$ 
and the three values $\mu=\mB$, $\mu=\mB/2$ and $\mu=2\mB$ of the coupling 
renormalization scale. The number of colours has been fixed to $n_c=3$ in 
this table, and the number of light flavours was varied from a very large 
negative value (equivalent to the large-$n_l$ limit) up to $n_l=10$. 
In columns 2 to 5 we show the ratios $c_n/\ac_{n}$, that correspond to an 
estimate of $N$ according to (\ref{eq:Ndet}) for finite $n$. In the last 
column we give $\Delta_{34}$. The $\pm$ numbers account for the change 
in $N$ due to the numerical uncertainty in the calculation of the 
exact four-loop conversion coefficient, which is about $0.1\%$ on the $n_l$ 
independent term for $\mu=\mu_m=m(\mu_m)$.

\begin{table}[t]
\begin{center}

{\small
\begin{tabular}{|c|| c c c c || c |}
 \hline
\multicolumn{6}{|c|}{$\mu/\mB=1$} \\
 \hline
 $\nl$ & $c_1/\ac_1$ & $c_2/\ac_2$ & $c_3/\ac_3$ & $c_4/\ac_4$ & $\Delta_{34}$ \\
 \hline
$-1000000	 $&$     0.6953 $&$     0.9624 $&$     0.9349 $&$     0.9714 $&$      0.038$\\
$-10	 $&$     0.4744 $&$     0.7152 $&$     0.6898 $&$     0.7005 \pm     0.0002 $&$      0.015 \pm      0.000$\\
$0	 $&$     0.4377 $&$     0.6357 $&$     0.6130 $&$     0.5977 \pm     0.0006 $&$      0.025 \pm      0.001$\\
$3	 $&$     0.3954 $&$     0.6150 $&$     0.5723 $&$     0.5370 \pm     0.0011 $&$      0.064 \pm      0.002$\\
$4	 $&$     0.3633 $&$     0.6120 $&$     0.5522 $&$     0.5056 \pm     0.0015 $&$      0.088 \pm      0.003$\\
$5	 $&$     0.3143 $&$     0.6119 $&$     0.5244 $&$     0.4616 \pm     0.0020 $&$      0.127 \pm      0.004$\\
$6	 $&$     0.2436 $&$     0.6089 $&$     0.4818 $&$     0.3942 \pm     0.0028 $&$      0.200 \pm      0.007$\\
$7	 $&$     0.1474 $&$     0.5378 $&$     0.4084 $&$     0.2786 \pm     0.0042 $&$      0.378 \pm      0.015$\\
$8	 $&$     0.0098 $&$     0.0379 $&$     0.2719 $&$     0.0564 \pm     0.0068 $&$      1.312 \pm      0.068$\\
$10	 $&$     0.2684 $&$    -0.0916 $&$    -0.1108 $&$    -1.7228 \pm     0.0271 $&$      1.758 \pm      0.004$\\
 \hline
\multicolumn{6}{|c|}{$\mu/\mB=0.5$} \\
 \hline
$-1000000	 $&$     1.3907 $&$     1.3554 $&$     0.6952 $&$     1.0773 $&$      0.431$\\
$-10	 $&$     0.9487 $&$     0.9410 $&$     0.6701 $&$     0.7110 \pm     0.0003 $&$      0.059 \pm      0.000$\\
$0	 $&$     0.8753 $&$     0.7907 $&$     0.6149 $&$     0.5807 \pm     0.0012 $&$      0.057 \pm      0.002$\\
$3	 $&$     0.7908 $&$     0.7343 $&$     0.5659 $&$     0.5030 \pm     0.0023 $&$      0.118 \pm      0.005$\\
$4	 $&$     0.7266 $&$     0.7159 $&$     0.5370 $&$     0.4631 \pm     0.0030 $&$      0.148 \pm      0.006$\\
$5	 $&$     0.6286 $&$     0.6975 $&$     0.4943 $&$     0.4078 \pm     0.0040 $&$      0.192 \pm      0.010$\\
$6	 $&$     0.4872 $&$     0.6704 $&$     0.4267 $&$     0.3243 \pm     0.0056 $&$      0.273 \pm      0.017$\\
$7	 $&$     0.2948 $&$     0.5640 $&$     0.3117 $&$     0.1845 \pm     0.0084 $&$      0.513 \pm      0.043$\\
$8	 $&$     0.0196 $&$     0.0370 $&$     0.1123 $&$    -0.0768 \pm     0.0135 $&$     10.676 \pm      5.676$\\
$10	 $&$     0.5367 $&$    -0.0621 $&$    -0.2877 $&$    -2.1014 \pm     0.0541 $&$      1.518 \pm      0.011$\\
 \hline
\multicolumn{6}{|c|}{$\mu/\mB=2$} \\
 \hline
$-1000000	 $&$     0.3477 $&$     0.6235 $&$     0.8631 $&$     0.9409 $&$      0.086$\\
$-10	 $&$     0.2372 $&$     0.4800 $&$     0.5883 $&$     0.6576 \pm     0.0001 $&$      0.111 \pm      0.000$\\
$0	 $&$     0.2188 $&$     0.4380 $&$     0.5217 $&$     0.5698 \pm     0.0003 $&$      0.088 \pm      0.001$\\
$3	 $&$     0.1977 $&$     0.4314 $&$     0.4947 $&$     0.5247 \pm     0.0006 $&$      0.059 \pm      0.001$\\
$4	 $&$     0.1817 $&$     0.4330 $&$     0.4831 $&$     0.5026 \pm     0.0007 $&$      0.040 \pm      0.001$\\
$5	 $&$     0.1572 $&$     0.4376 $&$     0.4681 $&$     0.4724 \pm     0.0010 $&$      0.009 \pm      0.002$\\
$6	 $&$     0.1218 $&$     0.4413 $&$     0.4452 $&$     0.4262 \pm     0.0014 $&$      0.044 \pm      0.003$\\
$7	 $&$     0.0737 $&$     0.3968 $&$     0.4038 $&$     0.3460 \pm     0.0021 $&$      0.154 \pm      0.006$\\
$8	 $&$     0.0049 $&$     0.0286 $&$     0.3177 $&$     0.1877 \pm     0.0034 $&$      0.515 \pm      0.017$\\
$10	 $&$     0.1342 $&$    -0.0761 $&$    -0.0083 $&$    -1.1238 \pm     0.0135 $&$      1.971 \pm      0.000$\\
 \hline
 \end{tabular}
}
\caption{The values of $N$ obtained from the coefficients of the perturbative
  expansion up to the fourth order for several values of $\nl$.  Three values
  of the renormalization scale are considered. }\label{tab:nc3vsnf}
  \end{center}
\end{table}

We first discuss the result for $\mu=\mB$.  For $\nl$ very large and negative 
the value of $N$ is close to the one predicted by~(\ref{eq:Nlargenl}). 
The value of $\Delta_{34}$ corresponds to a $4\%$ deviation of the 
third order coefficient from the asymptotic result, which is indeed the 
case, and the fourth-order value is already much 
closer.\footnote{We may note that the contribution 
from sub-leading renormalon poles to $c_n$ is of order $1/2^n$ 
relative to the leading one, but there is a further suppression for the 
case at hand due to a 
small numerical coefficient, at least in the large-$n_l$ limit, 
see~\cite{Beneke:1998ui}.} As $\nl$ increases, the value of $N$
decreases, reaching $0.506(2)$ and $0.462(2)$ for $\nl=4$ and 5, 
respectively, with a 9 and 13\%{} variation when going from the third to 
the fourth order coefficient. As $\nl$ increases, $\Delta_{34}$ also
increases, so that for $\nl$ above 7 the $N$ values obtained from the
third and fourth order coefficients differ by factors of order 1.
This behaviour is not unexpected: by increasing the number of light flavours
the first coefficient of the $\beta$ function, $b_0$, decreases (it vanishes 
for $\nl=33/2$), hence the renormalon dominance is delayed to higher 
orders. We shall comment further on the $n_l$ dependence below.

\begin{figure}[t]
  \begin{center}
    \includegraphics[width=0.6\textwidth]{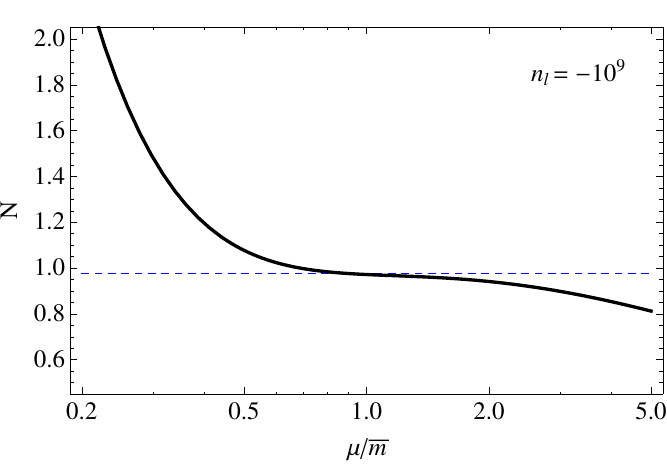}
  \end{center}
  \vspace*{-0.4cm}
  \caption{\label{fig::Nlargenl}
    The normalization $N$ as a function of $\mu/\overline{m}$ varied by a
    factor five around the central scale. The dashed line shows the 
    exact value $4 e^{5/6}/(3\pi) = 0.97656 ...$.}
\end{figure}

When considering different choices of the renormalization scale, we
see that the $\mu=\mB/2$ case leads to larger variations than  $\mu=2 \mB$.
The large $\nl$ limit yields a value that is about 10\%{} higher than the 
exact result but the associated value $\Delta_{34}\approx 40\%$ is 
also large, indicating that the series is not as close to the asymptotic 
regime as for $\mu=\mB$. For the interesting cases $\nl=4$ and $\nl=5$, 
$\Delta_{34}$ is also more than a factor of two larger than for $\mu=\mB$. 
Again, this behaviour is not unexpected. The coefficients $c_n$ depend 
only on logarithms of $\mu/m$ up to the $(n-1)$th power. 
Eq.~(\ref{eq:cnasymp}) shows that these logarithms must asymptotically 
exponentiate to $\mu/m$, which clearly happens less efficiently at finite 
order when $\ln(\mu/m)$ is larger. Hence we expect the best approximation 
to the asymptotic behaviour to occur when $\mu\approx \mB$. 
Fig.~\ref{fig::Nlargenl} shows that this is indeed the case for 
large $-n_l$. It further shows a plateau around $\mu\approx \mB$ 
and a more rapid departure from the exact result for $\mu$ smaller 
then $\mB$ than for larger $\mu$, as also seen in table \ref{tab:nc3vsnf}.

\begin{figure}[t]
  \begin{center}
  \includegraphics[width=0.5\textwidth]{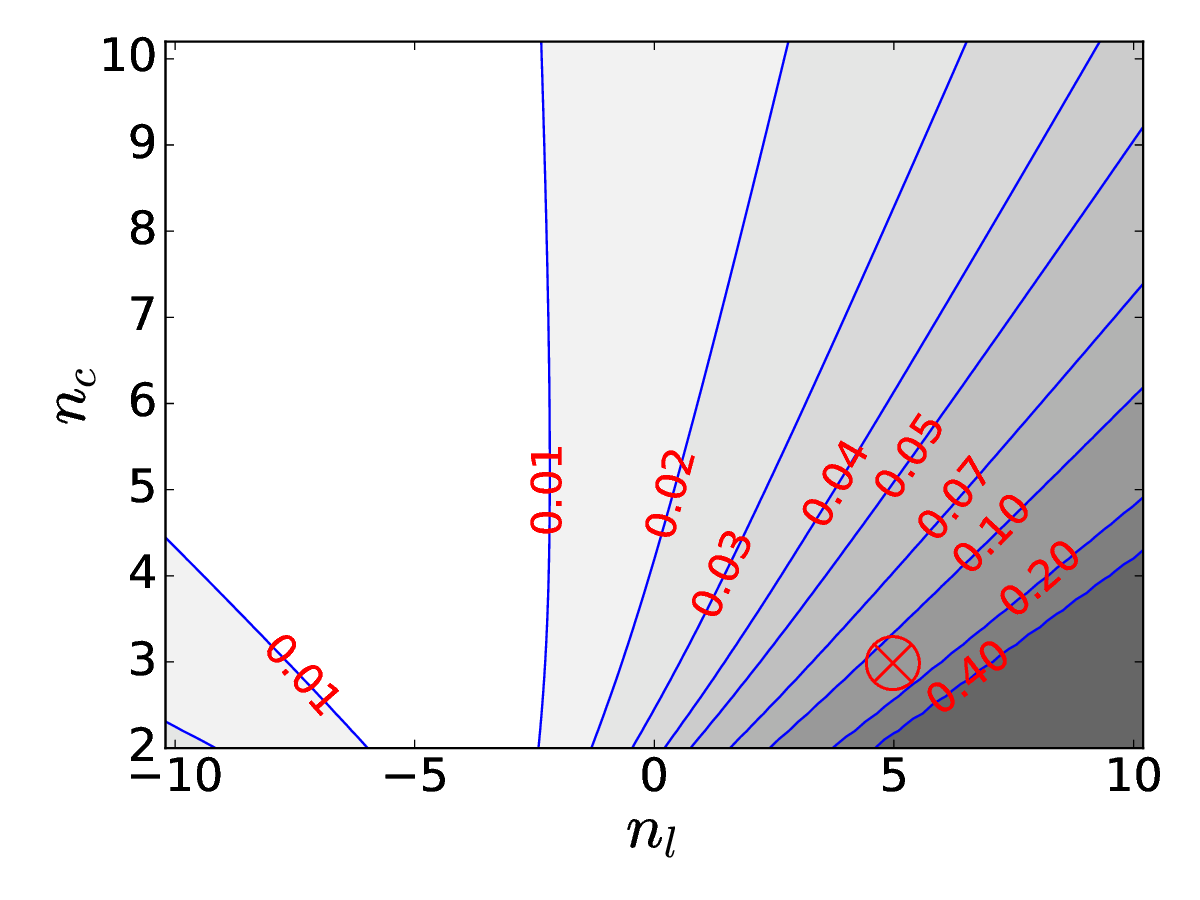}
  \end{center}
  \vspace*{-0.4cm}
\caption{\label{fig:ncnf1}
$\Delta_{34}$ as a function of $\nc$ and $\nl$,  for $\mu=\mu_m=\mB$.
The cross corresponds to the case relevant for top, i.e. $\nc=3$ and 
$\nl=5$.}
\end{figure}

We also determine the normalization $N$ for different values of $n_c$ 
and show the result for $\Delta_{34}$ in fig.~\ref{fig:ncnf1}. We 
generically find $\Delta_{34}<0.1$ except in regions where $b_0$ is small, 
where we do not expect our method to work. 
Fig.~\ref{fig:ncnf1} therefore demonstrates that 
the exact four-loop coefficient indeed matches the asymptotic 
formula~(\ref{eq:beneke2}) in the expected range of $\nc$ and $\nl$ 
values, comprising those of physical interest. 

\begin{figure}[t]
  \begin{center}
  \includegraphics[width=0.45\textwidth]{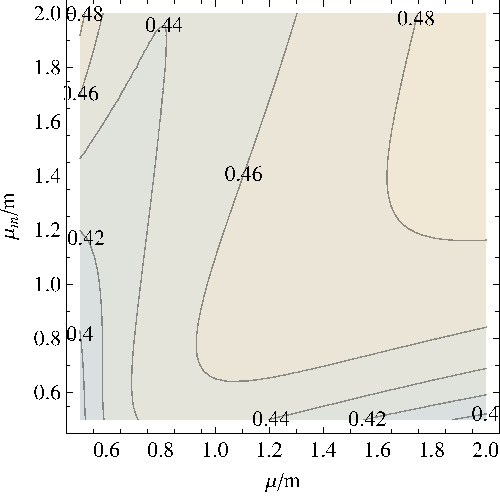}
  \end{center}
  \vspace*{-0.4cm}
  \caption{\label{fig::norm_mur_muf2}
  The normalization $N$ as a function of $\mur/m(\mum)$ and $\mum/m(\mum)$.}
\end{figure}

For the following a reliable determination of $N$ and an estimate of 
its error is particularly important for $n_c=3$, $n_l=5$, corresponding 
to the case of the top quark. We determine the error by varying the 
two renormalization scales independently, that is we vary 
$\mur/m(\mum)$ and $\mum/m(\mum)$ independently between 0.5 and 2, 
compute $N$ from $c_4/\ac_4$ as above, 
and determine the error on $N$ from the maximal variation. The dependence 
of $N$ on the two scale ratios is shown in fig.~\ref{fig::norm_mur_muf2}. 
With this definition our error estimate on $N$ neither 
depends on the value of the heavy quark mass nor the one of the strong 
coupling. We find 
\begin{equation}
\label{eq:Nestimate}
N= 0.4616{}^{+0.027}_{-0.070}\,(\mur \mbox{ and } \mum) \pm 0.002\,(c_4)\,.
\end{equation}
As a further check we note that when the subleading term $s_2$ 
($s_1$ and $s_2$) is removed in (\ref{eq:ctildenasymp}), the 
central value changes very little to 0.4573 (0.4584).\footnote{
Using the five-loop beta-function coefficient 
from \cite{Baikov:2016tgj}, which appeared after this analysis was 
finished, allows us to compute the 
next sub-asymptotic term $s_3$ in~(\ref{eq:ctildenasymp}) (see appendix). 
We find that $N$ changes by a negligible 
amount to 0.4606.}

A similar method to determine the normalization of the leading pole 
mass renormalon, albeit without variations of $\mu_m$ and $n_c$, has 
already been used in \cite{Ayala:2014yxa}. More precisely, instead of 
the four-loop pole mass considered here the three-loop static potential 
was employed to arrive at the best estimate, based on the fact that the 
pole mass and static potential leading renormalon normalizations are 
rigorously related by a factor of $-1/2$.  Their values are indeed in 
good agreement with ours, though deteriorating with increasing $n_l$. 
The approach to the exact value for large negative $n_l$ was 
also observed in \cite{Ayala:2014yxa}.

\begin{figure}[b]
  \begin{center}
    \includegraphics[width=0.6\textwidth]{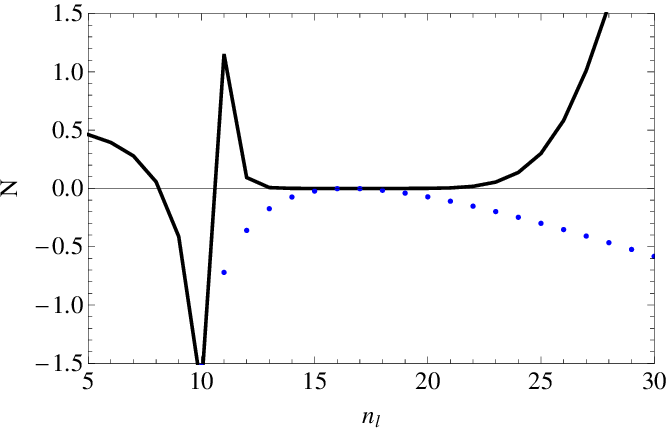}
  \end{center}
  \vspace*{-0.4cm}
  \caption{\label{fig:nlplateau}
    The normalization $N$ (for $n_c=3$, $\mu=\mu_m=\mB$) as a function 
    of $n_l$ (black). The blue dots show $1/b$.}
\end{figure}

The authors of~\cite{Ayala:2014yxa} also determined the normalization 
$N$ as a function of $n_l$ and noted that it tends to zero in the 
range $n_l=12\ldots 23$ close to the conformal window. We confirm 
this behaviour in our analysis, see Figure~\ref{fig:nlplateau}. 
To understand why the normalization of the leading renormalon is forced to 
be small in this $n_l$ region, we look at the explicit expression of 
for $c_n^{(\rm as)}$ from (\ref{eq:cnasymp}) for $n=4$,
\begin{equation}
c_4^{(\rm as)} =
(2 b_0)^3\, (1+b)(2+b)(3+b) \left(1+\frac{s_1}{3+b}+\frac{s_2}{(3+b)(2+b)}
+\cdots \right)\,.
\end{equation}
The region $n_l=12\ldots 23$ is approximately centred around the value 
of $n_l$, where $b_0$ vanishes, hence $b=b_1/(2b_0^2)$ becomes large. 
As soon as $b\gg n_0$, where $n_0$ is the order from which $N$ is 
determined (here $n_0=4$), the individual terms in the above expression 
behave as 
\begin{equation}
c_4^{(\rm as)} = (2 b_0)^{n_0} \left(\frac{b_1}{2 b_0^2}\right)^{\!n_0} 
\left(1+\frac{s_1}{b}+\frac{s_2}{b^2}+\cdots\right)
\sim \frac{1}{(2 b_0)^{n_0}} 
\left(1+\frac{\#}{b_0^2}+\frac{\#}{b_0^4}+\cdots\right),
\end{equation}
from which we conclude a) that $c_4^{(\rm as)} \sim 1/(2 b_0)^{n_0}$ 
becomes very large, hence $N$ {\em must$\,$} become small to fit the 
given value of the exact four-loop coefficient $c_4$, and b) the series 
of sub-leading asymptotic terms $s_1$, $s_2$, etc. breaks down, 
hence the extracted value of $N$ is completely unreliable. The 
smallness of $N$ is therefore a technical artifact of the method, which 
ceases to be valid when $b$ becomes large compared to $n_0$, and the 
question whether $N$ is small in the conformal window cannot be answered. 
In fact, while small $b_0$ makes renormalon behaviour less relevant to 
low orders due to the diminished $(2 b_0)^n$ factor, 
there seems  to be no reason why the normalization $N$ 
should vanish when the theory becomes conformal non-perturbatively.


\section{\boldmath 
The $\mpole$ -- $\mB$ conversion factor to all orders 
and the ultimate top pole mass uncertainty}
\label{sec:extrapolation}

In the following we use two methods to estimate the remainder of the 
mass conversion relation beyond the exactly known four-loop accuracy 
and to estimate the intrinsic ambiguity of summing the assumed  
asymptotic expansion. The first relies on truncation of the expansion 
and an estimate of the minimal term. The second on Borel summation. 
We restrict ourselves to the case of the top quark mass ($n_c=3$, $n_l=5$) 
and choose $\mu=\mu_m=\mB$.

We begin by writing 
\begin{equation}\label{eq:polemassn}
\mpole(n)=\mB\left(1+\sum_{k=1}^n c_k \as^{k}\right)\,,
\end{equation}
where the coefficients are the exact ones up to the fourth order in 
\as{}, and determined from the asymptotic formula (\ref{eq:ctildenasymp}) 
(with normalization fitted to the fourth order term) for the terms of order 
5 and higher. We would like to define the best value of \mpole{} as the 
value at which its increment with $n$ is minimal. More precisely, we define
\begin{equation}
\Delta(n+1/2) = \mpole(n+1)-\mpole(n)\,,
\end{equation}
which is a decreasing function of $n$ up to a certain value $n_0$ 
beyond which it begins to increase due to the renormalon divergence of 
the series expansion.
By interpolating $\Delta$ with a quadratic form in the three points 
$n_0-1/2$, $n_0+1/2$, $n_0+3/2$, we find its minimum at (generally 
non-integer)
\begin{equation}
n_{\rm min}=n_0+1/2-\frac{\Delta(n_0+3/2)-\Delta(n_0-1/2)}
{2(\Delta(n_0+3/2)+\Delta(n_0-1/2)-2\Delta(n_0+1/2))}.
\end{equation}
By interpolating linearly the value of $\mpole(n_{\rm min})$ between 
$n_0$ and $n_0+1$ we get
\begin{equation}\label{eq:cerntralvalue}
\mpole^{\rm c}=\frac{\mpole(n_0) (\Delta(n_0+3/2)-\Delta(n_0+1/2))
+\mpole(n_0+1)(\Delta(n_0-1/2)-\Delta(n_0+1/2))}
{\Delta(n_0+3/2)+\Delta(n_0-1/2)-2\Delta(n_0+1/2)}
\end{equation}
as the best value of the pole mass. We note that with this prescription, 
if $\Delta(n_0-1/2)=\Delta(n_0+3/2)$, then $\mpole^{\rm c}$ corresponds 
to $(\mpole(n_0)+\mpole(n_0+1))/2$, as one would intuitively expect, 
while for $\Delta(n_0-1/2)\gg\Delta(n_0+3/2)$ 
($\Delta(n_0-1/2)\ll\Delta(n_0+3/2)$), we obtain $\mpole(n_0+1)$ 
($\mpole(n_0)$).

\begin{table}[t]
  \begin{center}
   \begin{tabular}{|c| c c |}
     \hline
     $j$ & $\acb_j$ & $\acb_j \as^j $ \\
     \hline
 $5$ & $    0.985499\times 10^2$ & $    0.001484$\\
 $6$ & $    0.641788\times 10^3$ & $    0.001049$\\
 $7$ & $    0.495994\times 10^4$ & $    0.000880$\\
 $8$ & $    0.443735\times 10^5$ & $    0.000854$\\
 $9$ & $    0.451072\times 10^6$ & $    0.000942$\\
 $10$ & $    0.513535\times 10^7$ & $    0.001164$\\
 $11$ & $    0.647283\times 10^8$ & $    0.001593$\\
 $12$ & $    0.894824\times 10^9$ & $    0.002390$\\
 $13$ & $    0.134620\times 10^{11}$ & $    0.003902$\\
 $14$ & $    0.218949\times 10^{12}$ & $    0.006888$\\
 $15$ & $    0.382818\times 10^{13}$ & $    0.013070$\\
      \hline
\end{tabular}
  \end{center}
  \caption{\label{tab:acbvalues}
    The coefficients $\acb_j$ above the fourth order. Their value
    multiplied by the corresponding power of 
    $\as = 0.108531$ is also reported. }
\end{table}

We now estimate the correction to the top pole mass due to terms of order
higher than four by
\begin{equation}\label{eq:delta5def}
\delta^{(5+)} \mpole = N \mur \,{\overline \sum_{k=5}} \acb_k \as^k(\mur)\,,
\end{equation}
where $\acb_j$ is defined in~(\ref{eq:ctildenasymp}), and the barred sum
represents the procedure we have just outlined for the evaluation of the
(divergent) sum. We report in table~\ref{tab:acbvalues} the values of 
$\acb_j$ beyond the fourth order term. Eq.~(\ref{eq:delta5def}) can be
easily computed for any value of $\as$ and $\mur$ and is well approximated 
by the second-order Taylor series around the reference value:
\begin{eqnarray}
  \delta^{(5+)} \mpole &=& N \mur \times 10^{-3} \left(3.604 + 14.69
    \left(\frac{\as(\mur)}{0.1085}-1\right) 
    +9.54 \left(\frac{\as(\mur)}{0.1085}-1\right)^2  \,\right).\quad
\end{eqnarray}
For typical values of $N\approx 0.5$ and $\mur\approx 160$~GeV the formula  
is accurate at the sub-MeV level for a $\pm 5\%$ variation of the 
strong coupling constant.

We now adopt the PDG value $\as(M_Z)=0.1181 \pm 0.0013$, and take
$\mur=\mB=163.508$~GeV for definiteness. With this input we find 
$\as(\mur)=0.108531$
for the (five flavour) strong coupling constant and $173.34$~GeV for 
the top pole mass using the four-loop conversion formula. 
From the values reported in the table and the value of $N$ 
given in~(\ref{eq:Nestimate}) we obtain for the series remainder 
\begin{eqnarray}
\label{remainder1}
    \delta^{(5+)} \mpole &=& 
     0.272^{+0.016}_{-0.041}\, (N) 
    \pm 0.001 \,(c_4) \pm 0.011 \, (\alpha_s) \pm 0.066\;
    \mbox{(ambiguity)} \;\mbox{GeV}\,,\qquad
\end{eqnarray}
where we show the error due to the uncertainty in the normalization $N$,  
the four-loop coefficient $c_4$, and $\alpha_s(M_Z)$. For the irreducible 
renormalon 
ambiguity we tentatively estimate the size of the first omitted term 
by the value of $\Delta(n_0-1/2)$. For the top mass conversion factor we 
find 
\begin{eqnarray}
  \mpole^{\rm c}/\mB &=& 1.06177 {}^{+0.00010}_{-0.00025} \,(N) \, 
  \pm 0.00001 \,(c_4)  \,  \pm 0.00087 \, (\alpha_s)\,
\nonumber\\
&&\pm \,0.00041 \;\mbox{(ambiguity)}\,.
\end{eqnarray}
We also computed the change of the conversion factor under variations 
of $\mu/\mB$ and $\mu_m/\mB$, simultaneously in the exact 
four-loop part and the remainder, accounting for the dependence of 
$N$ on $\mu$ and $\mu_m$ (fig.~\ref{fig::norm_mur_muf2}). 
This leads to ${}^{+0.00025}_{-0.00041}$, 
which we do not include above, since it is strongly correlated 
with the uncertainty of the same order from $N$ alone. 

In the second method we first compute the Borel transform of the asymptotic 
series coefficients $\acb$ in (\ref{eq:cnasymp}), which gives
\begin{equation}
B[\acb](t) = \frac{1}{(1-2b_0 t)^{1+b}} + 
\frac{s_1}{b}\,\frac{1}{(1-2b_0 t)^{b}} + 
\frac{s_2}{b (b-1)}\,\frac{1}{(1-2b_0 t)^{-1+b}} + \ldots,
\end{equation}
and then the Borel sum 
\begin{equation}
BS[\acb](\alpha_s) = \int_0^\infty dt\,e^{-t/\alpha_s}\,B[\acb](t)\,.
\end{equation}
Since the series is not Borel-summable due to the IR renormalon singularity 
at $t=1/(2 b_0)$, we define the sum as the principal value and estimate the 
ambiguity as the imaginary part of the integral when the contour is deformed 
into the upper complex plane, divided by Pi. This procedure is known to 
usually give a reliable estimate~\cite{Beneke:1998ui}, 
close to the sum to the minimal term 
and the estimate of the summation ambiguity by the smallest term 
in the series. The Borel sum can easily be computed analytically, 
since (with the contour deformed into the upper complex plane)
\begin{equation}
\int_0^\infty dt\,e^{-t/\alpha_s}\,\frac{1}{(1-2b_0 t)^\gamma} 
= \frac{\alpha_s}{(-2 b_0\alpha_s)^\gamma}\,e^{-1/(2b_0\alpha_s)}\,
\Gamma(1-\gamma,-1/(2 b_0\alpha_s))\,,
\end{equation}
where $\Gamma(a,z)$ denotes the incomplete Gamma function. The remainder 
of the mass conversion formula is obtained by subtracting the 
first four coefficients, resulting in 
\begin{equation}
\label{borelremainder}
 \delta^{(5+)} \mpole = N\mu
 \left(BS[\acb](\alpha_s(\mu)) - \sum_{k=1}^4 \acb_k \alpha_s(\mu)^k\right)\,.
\end{equation}
With parameter input as above, we find 
\begin{eqnarray}
  \label{remainder2}
  \delta^{(5+)} \mpole &=& 0.250 {}^{+0.015}_{-0.038} \, (N) 
  \pm 0.001 \, (c_4) \pm 0.010 \, (\alpha_s) \pm 0.071 
  \;\mbox{(ambiguity)}\;\mbox{GeV}\,, \qquad
\end{eqnarray}
which is close to the result (\ref{remainder1})
from the previous method. For any value of $\as$ and $\mur$ the result 
can again be determined accurately in the phenomenologically relevant  
region according to the fit formula
\begin{eqnarray}
  \delta^{(5+)} \mpole &=& N \mur \times 10^{-3}  \left(3.315 + 12.71
    \left(\frac{\as(\mu)}{0.1085}-1\right) 
    + 4.55 \left(\frac{\as(\mu)}{0.1085}-1\right)^2  \,\right).\qquad
\end{eqnarray}
For the top mass conversion factor itself, we find 
\begin{eqnarray}
  \mpole^{\rm c}/\mB &=& 1.06164{}^{+0.00009}_{-0.00023}  \,(N) 
  \pm 0.00001 \,(c_4) \,  \pm 0.00086 \, (\alpha_s)\,
\nonumber\\
&& \pm \,0.00043
  \;\mbox{(ambiguity)}.\qquad
\label{eq:convfinal}
\end{eqnarray}
In this case, the scale variation is ${}^{+0.00013}_{-0.00028}$. 

The ultimate uncertainty on the top quark pole mass, 
which we identify with the ambiguity  
of about 70 MeV, is smaller than estimates from the 
large-$n_l$ limit, because the normalization $N$ is smaller. 
We also note that dividing the imaginary part of the Borel integral by
Pi to obtain the ambiguity is a convention that has proven reliable in
contexts where the quantity in question is amenable of a
non-perturbative definition~\cite{Beneke:1998ui}.
This is not the case for the pole mass, so that we cannot ask how well
the divergent series approximates the exact, non-perturbative result. The
point is rather that the pole mass can in principle be used as a
reasonable perturbative reference parameter, as long as computing
additional orders does not require increasingly larger shifts in the
reference value. The dividing-by-Pi convention therefore appears 
reasonable, since, if the imaginary part of the Borel transform was 
instead used to estimate the ambiguity, it would be almost as large as 
the known four-loop term, where the series is clearly still in the regime of
decreasing terms. We observe that, in any case, even if the ambiguity
were taken to be the imaginary part of the Borel integral itself, the
resulting estimate of would still be significantly below the
uncertainty that can conceivably be achieved at hadron colliders.


\section{Internal bottom and charm mass effect}
\label{sec:masseffect}

The analysis assumed up to now that the five lighter quarks are massless. 
Since the typical loop momentum at order $\alpha_s^{n+1}$ is of order 
$m_t e^{-n}$ in the regime where the series is dominated by the leading
renormalon divergence, we expect internal quark mass effects from the 
bottom and charm quark to become more important in higher orders. Furthermore, 
the minimal term is attained when the typical loop momentum is of order 
$\LambdaQCD$, hence the ambiguity should be determined by 
$\Lambda$-parameter $\LambdaQCD^{(3)}$ in the three-flavour scheme, 
excluding the bottom and charm quark. In this section we estimate the 
effect of the finite bottom and charm quark mass on the top mass conversion 
factor and the ultimate uncertainty.

The decoupling of internal quark loops from quarks with masses 
$m_q\gg \LambdaQCD$ in the renormalon asymptotic behaviour was studied 
analytically and numerically in the large-$n_l$ limit \cite{Ball:1995ni}. 
The analysis showed that the asymptotic behaviour of the series in a 
theory with $n_l$ quarks of which $n_m$ are massive, approaches the series 
of the theory with $n_l-n_m$ massless quarks when both are expressed in 
terms of the $\overline{\rm MS}$ coupling $\alpha_s^{(n_l-n_m)}(m_t)$ in 
the $n_l-n_m$ flavour scheme.\footnote{Note that (4.14) 
in \cite{Ball:1995ni} does not apply term by term, but only as a 
transformation of the entire series. Term by term the approximation 
holds, if the right-hand side 
of (4.14) is multiplied by the factor 
\[
\exp\left(\frac{1}{12\pi \beta_0^{(3)}}\ln\frac{m_b^2}{m_c^2}\right)/\left(1-
\frac{\alpha_s^{(3)}}{6\pi}\ln\frac{m_b^2}{m_c^2}\right)^{n+1},
\]
which follows 
from (4.16) in \cite{Ball:1995ni}. Here we put $\beta_0^{(3)}$ into 
the exponent rather than $\beta_0^{(4)}$ as in (4.16), since in the presence 
of a massive quark, the leading singularity is slightly shifted to 
$u=1/2\times 
\beta_0^{(4)}/\beta_0^{(3)}$ when $u$ is defined as $-\beta_0^{(4)} t$.}
Based on this observation it has been argued  
\cite{Ayala:2014yxa} that the bottom mass conversion factor should be 
expressed in terms of $\alpha_s^{(3)}(m_b)$ rather than 
the four-flavour coupling $\alpha_s^{(4)}(m_b)$. For the two- and three-loop 
coefficients, for which the mass dependence is known 
\cite{Gray:1990yh,Bekavac:2007tk}, it was shown that this substitution indeed renders the charm mass effect almost negligible. 

This procedure does not work for top, however, since the masses of 
the bottom and charm quark are too small in relation to $m_t$ to express 
the {\em entire} series in terms of the four- or three-flavour coupling. 
Instead, we switch from the five- to the four-flavour scheme at the 
order, where the typical internal loop momentum is of order $m_b$, 
which is ${\cal O}(\alpha_s^5)$, and from the four- to the three-flavour 
scheme at ${\cal O}(\alpha_s^6)$. Since the mass effect is not known 
for $c_4$ at the four-loop order, and since $c_n$ beyond the four-loop 
order can only be estimated assuming dominance of the first renormalon 
(as done above), this implies the following 
procedure: (a) at two- and three-loops we include the known mass 
dependence, but $c_4$ is approximated by the massless value. For given 
top $\overline{\rm MS}$ mass, this increases the top pole mass by 
11 (2-loop) + 16 (3-loop) MeV, adopting $\overline{m}_b=4.2$~GeV 
and $\overline{m}_c=1.3$~GeV. Since the $c_n$ increase as $n_l$ 
decreases, the mass effect is also expected to be positive in higher orders. 
Hence approximating $c_4$ by its massless value {\em underestimates} the 
mass effect. (b) At five-loop, we use $c_5^{(\rm as)} [\alpha_s^{(4)}(m_t)]^5$ 
with $c_5^{(\rm as)}$ determined as described in 
sect.~\ref{sec:renormalon}, but 
with the normalization $N_m=0.5056$ and beta-function coefficients for 
the four-flavour theory, $n_l=4$. (c) Beyond five loops, the remainder 
and the ambiguity is calculated according to (\ref{borelremainder}) 
(with obvious modification, since we sum the terms from six rather 
than five loops), but with the three-flavour scheme coupling 
$\alpha_s^{(3)}(m_t)$ and normalization $N_m=0.5370$. Since the bottom 
and charm quarks are not yet completely decoupled at the five- to 
seven-loop order, and since an extra quark flavour decreases the $c_n$, 
we expect that (b) and (c) {\em overestimate} the mass effect, since 
the approximation assumes that bottom and charm are already decoupled 
completely. The sum of (b) and (c) adds another 53 MeV to the top pole 
mass, such that the total mass effect is estimated to be 80~MeV. 
Since the bottom is neither heavy enough to be decoupled in low orders, 
nor light enough to be ignored, where in both cases a massless approximation 
can be justified, there is an inherent uncertainty in the above estimate. 
However, as argued above, the errors in the approximations are expected 
to go in opposite directions, hence we consider $(80\pm 30)$~MeV a 
conservative estimate of the internal bottom and charm quark mass effect 
on the top pole mass. The 30~MeV error estimate arises from an estimate 
of the neglected mass effect on $c_4$ by extrapolation from the known 
lower orders. We have also checked that the approximation described here 
works well in models for the series inspired by the large-$n_l$ limit. 

Including the internal mass effect into the massless results
(\ref{remainder2}) and (\ref{eq:convfinal}), we obtain for the series 
remainder from the five-loop order
\begin{eqnarray}
  \label{remainder3}
  \delta^{(5+)} \mpole &=& 0.304 {}^{+0.012}_{-0.063} \, (N) 
  \pm 0.030 \, (m_{b,c}) \pm 0.009 \, (\alpha_s) \pm 0.108 
  \;\mbox{(ambiguity)}\;\mbox{GeV}\,, \qquad
\end{eqnarray}
where we now dropped the negligible uncertainty from the massless 
four-loop coefficient $c_4$. Apart from the shift of the value 
of $\delta^{(5+)}\mpole$ the ambiguity has increased to 108~MeV, which is 
mainly due to the fact that $\LambdaQCD^{(3)}$ is larger than 
$\LambdaQCD^{(5)}$. Note that the ambiguity is independent of 
the precise value of the bottom and charm mass, as long as 
$m_b, m_c \gg \LambdaQCD$. This also implies that it is the same 
for any heavy quark, including the bottom quark, since it depends 
only on the infrared properties of the theory, which is QCD with three 
approximately massless flavours.

For the top mass conversion factor itself, we find 
\begin{eqnarray}
  \mpole^{\rm c}/\mB &=& 1.06213{}^{+0.00007}_{-0.00038}  \,(N) 
  \pm 0.00018 \,(m_{b,c}) \,  \pm 0.00086 \, (\alpha_s)\,
\nonumber\\
&& \pm \,0.00066
  \;\mbox{(ambiguity)}.\qquad
\label{eq:convfinal2}
\end{eqnarray}
The scale variation remains as for (\ref{eq:convfinal}).
We adopt (\ref{remainder3}) and (\ref{eq:convfinal2}) as our final results. 
Given the $\overline{\rm MS}$ mass, the top quark pole mass is determined by 
this relation with an accuracy of 1.1 per mil, half of which 
is due to the irreducible uncertainty of the relation itself. 


\section{Conclusions}\label{sec:conclusions}

We employed the four-loop coefficient in the pole-$\overline{\rm MS}$ 
quark mass relation, which has recently become 
available~\cite{Marquard:2015qpa}, and knowledge of the leading 
asymptotic behaviour of the series expansion of the 
mass conversion factor~\cite{Beneke:1994rs} to estimate 
the remainder of the series from terms above the four-loop order 
and the intrinsic ambiguity due to the asymptotic nature of the 
series. For the case of the top quark we find about $300~$MeV for 
the former, including an estimate of the effect of the internal bottom and 
charm quark mass, and $110~$MeV for the ambiguity, which also represents 
the ultimate precision that can be obtained for the pole mass. 
The ambiguity of $110~$MeV is far below the accuracy that can conceivably 
be achieved at the Large Hadron Collider, but larger than the one 
foreseen in theoretical and experimental studies 
\cite{Beneke:2015kwa,Simon:2016htt} of a scan of the top pair production 
threshold at a high-energy $e^+ e^-$ collider. In this case 
the pole mass ceases to be a useful concept and other mass definitions 
must be employed.

\subsubsection*{Acknowledgements}

This work is supported by the BMBF grants 05H15WOCAA (MB) and 
05H15VKCCA (MS). PM was supported in part by the EU
Network HIGGSTOOLS PITN-GA-2012-316704. MB thanks the Kavli Institute for 
Theoretical Physics, Santa Barbara, for hospitality while this work was 
completed.


\appendix
\section{Summary of formulae}
\label{sec:appendixA}

In this Appendix, in order to make contact with the notation
of \cite{Beneke:1994rs,Beneke:1998ui}, we define
the QCD beta-function as
\begin{equation}
\beta(\alpha_s) = \mu^2\frac{\partial \alpha_s(\mu)}{\partial \mu^2} 
= \beta_0 \alpha_s^2+\beta_1\alpha_s^3 +\ldots,
\end{equation}
With this convention $\beta_0 = -(11 \nc/3- 2 \nl/3)/(4\pi)$, while in 
the main text we used $b_i=-\beta_i>0$ (for small $n_l$).We adopt 
the $\overline{\rm MS}$ scheme with $\nl$ massless quark flavours. (The 
heavy quark whose mass is considered here is decoupled.)
The constants that appear in 
(\ref{eq:ctildenasymp}) are given by \cite{Beneke:1994rs,Beneke:1998ui}
$b=-\beta_1/(2\beta_0^2)$ and 
\begin{eqnarray}
\label{eq:ctildenasympparams}
s_1 &=& \left(-\frac{1}{2\beta_0}\right)\left(-\frac{\beta_1^2}{2
\beta_0^3}+\frac{\beta_2}{2\beta_0^2}\right),
\\[0.2cm]
s_2 &=& \left(-\frac{1}{2\beta_0}\right)^2\left(\frac{\beta_1^4}{8
\beta_0^6}+\frac{\beta_1^3}{4\beta_0^4}-\frac{\beta_1^2\beta_2}{4
\beta_0^5}-\frac{\beta_1\beta_2}{2\beta_0^3}+\frac{\beta_2^2}{8
\beta_0^4}+\frac{\beta_3}{4\beta_0^2}\right),
\\[0.2cm]
s_3 &=& 
\left(-\frac{1}{2\beta_0}\right)^3\bigg(
-\frac{\beta_1^6}{48 \beta_0^9}
-\frac{\beta_1^5}{8\beta_0^7}
-\frac{\beta_1^4}{6 \beta_0^5}
+\frac{\beta_1^4\beta_2}{16 \beta_0^8}
+\frac{3 \beta_1^3 \beta_2}{8 \beta_0^6}
+\frac{\beta_1^2 \beta_2}{2 \beta_0^4}
-\frac{\beta_1^2 \beta_2^2}{16 \beta_0^7}
\nonumber\\
&& 
-\frac{\beta_1^2 \beta_3}{8 \beta_0^5}
-\frac{\beta_1 \beta_2^2}{4 \beta_0^5}
-\frac{\beta_1 \beta_3}{3 \beta_0^3}
+\frac{\beta_2^3}{48\beta_0^6}
-\frac{\beta_2^2}{6\beta_0^3}
+\frac{\beta_2 \beta_3}{8 \beta_0^4}
+\frac{\beta_4}{6 \beta_0^2}
\bigg)\,.
\end{eqnarray}
Note that we have corrected some misprints in the expression for 
$b$ and $s_2$ given in \cite{Beneke:1998ui} (eqs.~(5.91) and (5.92)) 
as already noted in \cite{Pineda:2001zq}. The result for $s_3$ was not 
given explicitly in \cite{Beneke:1998ui}. 

\providecommand{\href}[2]{#2}\begingroup\raggedright\endgroup

\end{document}